# Real-time geopotentiometry with synchronously linked optical lattice clocks

Tetsushi Takano[1,2], Masao Takamoto[2,3,4], Ichiro Ushijima[2,3,4], Noriaki Ohmae[1,2,3], Tomoya Akatsuka[2,3,4], Atsushi Yamaguchi[2,3,4], Yuki Kuroishi[5], Hiroshi Munekane[5], Basara Miyahara[5] & Hidetoshi Katori[1,2,3,4]

[1]*Department of Applied Physics, Graduate School of Engineering, The University of Tokyo, Bunkyo-ku, Tokyo 113-8656, Japan.*

[2]*Innovative Space-Time Project, ERATO, Japan Science and Technology Agency, Bunkyo-ku, Tokyo 113-8656, Japan.*

[3]*Quantum Metrology Laboratory, RIKEN, Wako-shi, Saitama 351-0198, Japan.*

[4]*RIKEN Center for Advanced Photonics, Wako-shi, Saitama 351-0198, Japan.*

[5]*Geospatial Information Authority of Japan, Tsukuba-shi, Ibaraki 305-0811, Japan.*

**According to the Einstein's theory of relativity, the passage of time changes in a gravitational field[1,2]. On earth, raising a clock by one centimetre increases its tick rate by 1.1 parts in $10^{18}$, enabling optical clocks[1,3,4] to perform precision geodesy[5]. Here, we demonstrate geopotentiometry by determining the height difference of master and slave clocks[4] separated by 15 km with uncertainty of 5 cm. The subharmonic of the master clock is delivered through a telecom fibre[6] to phase-lock and synchronously interrogate[7] the slave clock. This protocol rejects laser noise in the comparison of two clocks, which improves the stability of measuring the gravitational red shift. Such phase-coherently operated clocks facilitate proposals for linking clocks[8,9] and interferometers[10]. Over half a year, 11**

**measurements determine the fractional frequency difference between the two clocks to be 1,652.9(5.9)×10$^{-18}$, or a height difference of 1,516(5) cm, consistent with an independent measurement by levelling and gravimetry. Our system is as a building block of an internet of clocks, consisting of a master and a number of slave clocks, which will provide "quantum benchmarks" that are height references with dynamic response.**

Stable disseminations of clock signals between metrology laboratories at the level of 10$^{-18}$ are being actively pursued with a future "redefinition of a second" in sight[11]. Optical frequency fibre links[12] using telecom fibres allow orders of magnitude better stabilities than those achieved with satellite links, and are therefore used for frequency comparisons requiring inaccuracies below 10$^{-16}$. Fibre links[13-15] over hundreds of kilometre are being developed worldwide and are used to compare distant optical clocks[16,17]; two optical clocks separated by 700 km have been compared, limited only by their clocks' uncertainties of 5×10$^{-17}$ (ref. 18) with no degradation due to the fibre link. In turn, such a fibre link is sensitive to the geopotential differences $\Delta\phi$ between the clocks via the gravitational red shift[2] $\Delta\nu/\nu_0 = \Delta\phi/c^2$ with $\nu_0$ the clock frequency, $c$ the speed of light. We envision a future internet of clocks consisting of "local clock networks" as shown in Fig. 1a. The clocks at the nodes of the fibre may take the place of conventional benchmarks maintained with time-consuming spirit levelling and gravimetry (see Methods).

We demonstrate a practical implementation of a "local clock network" that makes full use of clocks' stabilities with minimal experimental resources. In optical lattice clocks that interrogate more than hundreds of atoms, the Dick-effect noise[19], originating from the frequency noise of the clock laser overwhelms the quantum projection noise and limits the clock instability[20]. Reducing the frequency noise of the lasers is the straightforward approach to improve the clock instability. However, the stabilities of mobile clock lasers[21] are often inferior to state-of-the-art laboratory-lasers, which are built with considerable experimental investments such as longer[22] and colder[23] cavities placed in an environment free from acoustic and seismic noises.

Figure 1b depicts such a situation. A "master clock" at RIKEN consists of cryogenic Sr optical lattice clocks that are interrogated by a "master laser" with an instability of $3\times10^{-16}$ at 1 s. The "slave clock" at The University of Tokyo (UTokyo) is interrogated by an inferior "slave laser" with an instability of $2\times10^{-15}$ at 1 s. The three clocks, RIKEN1, RIKEN2, and UT, operate on the $^1S_0 - {}^3P_0$ transition of $^{87}$Sr at $\nu_0 \approx 429.2$ THz (698 nm), and are connected by a 30-km-long telecom fibre. The "master clock" delivers a subharmonic of the clock signal $\nu_0/2 \approx 214.6$ THz (1397 nm) to the slave clock through the fibre without using any frequency combs[13,16,17]. This configuration significantly simplifies conventional clock links[16-18] and allows long-term reliable operation.

Cryogenic clocks[4] suppress the uncertainty due to the blackbody radiation shift to below $1\times10^{-18}$ by surrounding lattice-trapped $^{87}$Sr atoms with light-absorbing copper enclosures at temperatures of about 100 K. After spin-polarizing the atoms, clock laser pulses with a duration of $T_i = 200$-$400$ ms interrogate the transition every $T_C = 1.5$ s. The "master clock" sends timing signals at 1.55 μm to the "slave clock" via the same fibre to synchronize the clock operation. The excitation probabilities of the clock transition are displayed in Figs. 2a-c. They steer the clock laser frequencies by applying $f_{\text{RIKEN1,2}}$ and $f_{\text{UT}}$ to frequency shifters (FS1, 2, and 3 in Fig. 1b). Phase locking the slave laser to the master laser (see Methods) reduces the scatter of the excitation probabilities of UT, as shown in Fig. 2a,b. Moreover, by operating the master and slave clocks synchronously, excitation probabilities correlate as shown in Fig. 2c.

The frequency difference between the clocks is directly obtained in this synchronous and phase-locked scheme from numerical frequency data applied to synthesizers (DDS1 and 3 in Fig. 1b), as the laser frequency is common to both the clocks. The clock shift is given by $\Delta\nu_{\text{UT-RIKEN1}}(t_j + \tau_d) = f_{\text{UT}}(t_j + \tau_d) - f_{\text{RIKEN1}}(t_j)$ for the $j$-th measurement sequence at $t_j = jT_C$ with $\tau_d \approx 0.16$ ms the transit time of the clock laser through the fibre. More importantly, this analysis rejects the correlated excitation noise originating from the Dick effect as a common noise[7] and improves the instability. Figure 2d summarizes the frequency instabilities $\Delta\nu_{\text{UT-RIKEN1}}/\nu_0$ with $T_i = 400$ ms, where each colour corresponds to the respective operation mode given in Figs. 2a-c. In particular, the instability measured for the phase-locked and synchronous operation decreases as interrogation times increase; $T_i = 200$ ms (green circles), 300 ms (blue circles), and 400 ms (red circles), in Fig. 2e. Dashed lines with corresponding colours show the Dick effect limited instability calculated for the fibre phase noise $h_L^{(T)} \approx 70$ (rad$^2 \cdot$ Hz)/km used in our experiments[6] (see Methods). This synchronous comparison should work more effectively for quiet fibres such as reported in the German link[24], which has $h_L^{(G)} \approx 1$ (rad$^2 \cdot$ Hz)/km, reducing the instability as shown by the coloured solid lines for the interrogation times above. Our observed instability of $6\times10^{-16}$ $(\tau/s)^{-1/2}$ obtained for $T_i = 400$ ms (red circles in Fig. 2e), which is severely degraded by the fibre noise, is already competitive with the "quantum coherence" spectroscopy demonstrated for co-located ions[8] of $4\times10^{-16}(\tau/s)^{-1/2}$. Moreover, the use of quiet fibre will further reduce the instability to $6\times10^{-17}(\tau/s)^{-1/2}$ with $N \approx 2{,}000$ atoms.

Figure 3a shows 11 measurements (blue circles) performed over 6 months. The frequency difference is determined to be $\Delta\nu_{\text{UT-RIKEN1}}/\nu_0 = -1{,}652.9(5.9)\times10^{-18}$ (blue shaded region), where the statistical and systematic uncertainties are $1.5\times10^{-18}$ and $5.7\times10^{-18}$, respectively. Table 1 summarizes the corrections and uncertainties for the "slave clock" $\nu_{\text{UT}}$ and the frequency difference $\Delta\nu_{\text{UT}-\text{RIKEN1}} = \nu_{\text{UT}} - \nu_{\text{RIKEN1}}$. The uncertainties of the frequency differences originating from the quadratic Zeeman shift, lattice light shift, clock light shift, servo error, and density shift are smaller than the square root of quadratic sum of the individual clocks' uncertainties due to the partial cancellation of the effects (see Methods). This frequency difference corresponds to a geopotential difference $\Delta\phi_{\text{UT}-\text{RIKEN1}}\left(=\frac{\Delta\nu}{\nu_0}c^2\right) = -148.55(53)$ m²/s². The result agrees with geodetically determined value $\Delta\phi_{\text{UT}-\text{RIKEN1}} = -148.14(6)$ m²/s² (red area), which is measured with spirit levelling and gravimetre by the Geospatial Information Authority of Japan in February, 2014. The right axis in Fig. 3a gives two results in terms of the geopotential difference.

The complete data set for the last experimental run in Fig. 3a (with $T_{\text{i}} = 300$ ms) is shown in Fig. 3b, which monitors the frequency difference for three days, with an uptime of 73 %. The two master clocks agree within the statistical uncertainty of $\Delta\nu_{\text{RIKEN2}-1}/\nu_0 = 1(1)_{\text{stat}} \times 10^{-18}$ (blue dots), demonstrating the reproducibility of these clocks. On the other hand, the master-slave comparison shows a gravitational red shift $\Delta\nu_{\text{UT}-\text{RIKEN1}}/\nu_0 = -1{,}653(7)_{\text{tot}} \times 10^{-18}$ (red dots), where the uncertainty includes systematic and statistical errors. Blue empty circles in Fig. 2e show the frequency instability of $\Delta\nu_{\text{UT}-\text{RIKEN1}}/\nu_0$, which reaches $6.4^{+2.0}_{-1.0} \times 10^{-18}$ after 6 hours, corresponding to a height uncertainty of 5.8 cm.

Figure 4 shows the calculated tidal perturbation $\phi_{\text{tidal}}(t)$ of the geopotential due to astronomical tides and ocean tidal loading for 15 days, including the days of the measurements (see Methods). The tidal perturbation (black line) is composed predominantly of semidiurnal and diurnal variations. The root-mean-square amplitudes of the perturbation are retained by about 70 % (red line) and about 92 % (blue line), for 6-hour and 3-hour averaging, respectively, suggesting 6 hour averaging is the characteristic time scale for clocks to reveal the tidal perturbation. While this tidal perturbation amounts to ≈ $6\times10^{-17}$ in terms of proper time passage (Fig. 4a), it is mostly common for clocks at UTokyo and RIKEN; therefore it cancels to ≈ $2\times10^{-19}$ in the clock comparison (Fig. 4b). This means that the clock shift directly measures the height difference $\Delta h = \frac{\Delta \nu_{\text{UT-RIKEN1}}}{\nu_0}\frac{c^2}{g} = -1{,}516(5)$ cm with $g = 9.798$ m/s$^2$ the gravitational acceleration. In contrast, as the phase of tidal perturbation changes, the potential difference $\Delta\phi_{\text{tidal}}(t)$ between Akune (one of tidal observation stations) and RIKEN, separated by 970 km and 9.4° of longitude, introduces peak to peak frequency variations $\Delta\phi_{\text{tidal}}(t)/c^2$ (the right axis in Fig. 4c), exceeding $1\times10^{-17}$. This is easily observable with a pair of master clocks with extended fibres (see Fig. 1a).

The instability $6\times10^{-16}(\tau/\text{s})^{-1/2}$ (see Fig. 2e) of our geopotentimetry with the "master and slave" configuration will be improved by an order of magnitude by applying a state-of-the-art "master laser" [22] to further extend the interrogation time or by employing less noisy fibres. The fibre noise can be also reduced by introducing repeaters. (see Methods) On the other hand, the systematic uncertainty of the clock comparison (see Table 1), which is dominated by the lattice-induced light shift, may be reduced to low $10^{-19}$ by employing an operational magic frequency[25] that nearly cancels the light shift perturbations arising from the multipolar and higher order polarizability effects. These strategies will allow mm-level clock-based altimetry with the time scale less than the tidal perturbation.

In summary, we have demonstrated cm-level clock-based altimetry using synchronously operated cryogenic optical lattice clocks, as a building block for an "internet of clocks". Such clock networks will be applicable to geodesy, fundamental physics and Earth sciences. For example, "quantum benchmarks" consisting of networked clocks will improve the realization of a long-range height reference frame by an order of magnitude, as clocks are not subjected to cumulative errors of several centimetres, levelling over a distance of 1,000 km (ref. 26). Such networks can unify the different height systems use in countries around the world. A global network of such clocks will also be able to search for topological dark matter[27]. Another possibility includes volcanic monitoring, consisting of a "local clock network" in an area within 100 km of an active volcano, by detecting the mass of ascending magma in a conduit as well as height changes induced by pressure changes at the chamber during a volcanic crisis.

## Methods

**Definition of height and its geodetic determination**

The commonly used notion of "height above sea level", or orthometric height in geodetic terms, is defined as the vertical distance to the geoid—the equi-geopotential surface that best fits to global mean sea level. Here, the geopotential of the earth is the sum of its gravitational potential and the potential of its centrifugal force due to its rotation. Orthometric height, simply referred to as "height", is determined by dividing the negative of the differential geopotential at a point with respect to the geoid by gravity. Conventionally, height difference between two points on the earth is precisely determined by geodetic techniques. Typically, a combination of spirit levelling and gravimetry determines height differences quite accurately over short distances[26] (with a precision less than 1 mm over a distance of 1 km), but its measurement uncertainty accumulates, predominately in proportion to the square root of the observation distances[26]. Moreover, the measurements can be time-consuming. For example, it takes approximately 10 years to complete the first-order levelling network of entire Japan, giving it inevitably poor time resolution. On the other hand, ellipsoidal height is geometrically defined as the normal distance to a reference rotational ellipsoid that best fits to global mean sea level. It can be measured with space geodetic positioning, such as Global Positioning System (GPS). The ellipsoidal height difference measured by GPS offers better time resolution (up to 1 Hz) but suffers from uncertainties introduced by the atmospheric delay in radio transmission (~ 1 cm after averaging over 24 h)[28].

**Correction of clock shifts**

Here, we focus on the estimation of the lattice shift, the servo error and the density shift, as the other effects are reported elsewhere[4].

**Lattice light shift**: Each optical lattice consists of two π-polarised counter-propagating laser beams, as described in ref. 4. The lattice frequencies of RIKEN and UT clocks are stabilised to 368,554,490.0(5) MHz and 368,554,490(3) MHz, respectively. The axial trap frequencies of atoms in the one-dimensional (1D) lattice are set to 40 kHz for all the clocks: RIKEN1, 2 and UT. The lattice light shift, including the multipolar and hyperpolarisability effect, is estimated to be $3.5(3.4)\times10^{-18}$ by measurements using RIKEN1 and RIKEN2. For the UT clock, the uncertainty caused by the fluctuation of the lattice frequency of 3 MHz is added to this value. For the frequency comparison budget, where common effects cancel out, the uncertainty is estimated to be $4.4\times10^{-18}$, which is the square root of the quadratic sum of the uncertainties arising from the travelling wave contamination[4] and from the lattice frequency fluctuation.

**Servo error**: Servo errors are estimated by calculating the mean value of the error signals of each measurement, which is obtained from the difference of the excitation fraction for right and left shoulders of the Rabi spectrum. Because the frequency noise of the master laser is shared by both clocks, the servo error of the two remote clocks correlate and are partly cancelled in calculating the frequency difference. We take the weighted average of the servo error for each measurement point and estimate the total servo error.

**Density shift**: UT and RIKEN1 clocks typically interrogate atoms $N_{\text{UT}} \approx 1{,}200$ and $N_{\text{RIKEN1}} \approx 450$, respectively. Applying the density shift of $0.9(4.2)\times10^{-18}$ measured for $N = 1{,}000$ atoms by interleaving the RIKEN1 and 2 clocks[4], we estimate the density shift for the UT clock to be $1.1(5.2)\times10^{-18}$ and RIKEN1 to be $0.4(1.9)\times10^{-18}$, assuming the shift and uncertainty to be proportional to $N$ for the same trap volume.

The density shifts can be partly cancelled out for the frequency comparison as a common effect. The uncertainty is estimated to be $3.3\times10^{-18}$ by taking the atom number difference $\Delta N = N_{\text{UT}} - N_{\text{RIKEN1}} \approx 750$.

**Transfer of a master laser stability and rejection of the Dick effects**

The master clock at RIKEN is equipped with a high-stability "master laser" $\nu_C^{\text{RIKEN}}$ stabilised to a 40-cm-long reference cavity, whereas at UTokyo, a "slave laser" $\nu_C^{\text{UT}}$ is stabilised to a 7.5-cm-long cavity. The stability of $\nu_C^{\text{RIKEN}}$ is estimated to be $3\times10^{-16}$ at 1 s based on the analysis of the clock excitation probabilities of RIKEN1 and 2, while the stability of $\nu_C^{\text{UT}}$ is measured to be $2\times10^{-15}$ at 1 s by the beat note with the transferred master laser from RIKEN as described below. These stabilities imply that the frequency noise of the master $\nu_C^{\text{RIKEN}}$ and the slave $\nu_C^{\text{UT}}$ lasers are by red and black dashed line, respectively, in Fig. 1c.

We employed the subharmonic $\nu_C^{\text{RIKEN}}/2$ to transfer the master laser stability to the slave, which is sent through the 30-km-long fibre with residual fibre noise shown by dashed orange line in Fig. 1c. Using the beat-note $f_{\text{beat}} = \nu_C^{\text{RIKEN}}/2 - \nu_C^{\text{UT}}/2$, the slave laser $\nu_C^{\text{UT}}$ is phase locked to $\nu_C^{\text{RIKEN}}$ with a frequency shifter 4 (FS4) for a bandwidth that minimizes the excess contamination of residual fibre noise. The slave laser is tightly locked to the transferred light for $f < 3$ Hz, where the Dick effect gives the largest contribution. For higher frequencies, $f > 15$ Hz, the servo gain is reduced to zero, as the fibre noise exceeds the frequency noise of $\nu_C^{\text{UT}}$. The blue line in Fig. 1c illustrates the frequency noise of $\nu_C^{\text{UT}}$ after the phase lock, which tracks the noise character of $\nu_C^{\text{RIKEN}}$ below 1 Hz. This enabled the slave laser at UTokyo to extend the interrogation time up to 600 ms.

Using the same laser frequency noise in both clocks rejects the common Dick effect, improving the frequency stability by 30% to $6\times10^{-16}/\sqrt{\tau/s}$, compared to the asynchronous case (see Fig. 2d). In linearized theory[6], this instability can be reduced to the red dashed line (see Fig. 2d), which gives the Dick effect for non-common fibre noise as shown by yellow shaded area in Fig. 1c (also see discussions in the following Methods). In contrast, the instability of $\nu_C^{UT}$ in stand-alone operation results in the Dick effect limited instability of $6 \times 10^{-15}/\sqrt{\tau/s}$, as shown by black circles in Fig. 2d.

**Frequency noise spectrum of clock lasers**

The power spectral density (PSD) of the frequency noise of the slave clock laser $\nu_C^{UT}$ after phase locking to the master laser is given by,

$$S_\nu^{UT}(f) = \left|\frac{G(f)}{1+G(f)}\right|^2 \left(S_\nu^{MASTER}(f) + S_\nu^{fibre}(f)\right) + \left|\frac{1}{1+G(f)}\right|^2 S_\nu^{SLAVE}(f),$$

where $S_\nu^{MASTER}(f)$ is PSD of the master laser $\nu_C^{RIKEN}$, $S_\nu^{SLAVE}(f)$ is that of the slave laser $\nu_C^{UT}$, $S_\nu^{fibre}(f)$ the delay-uncompensated fibre noise, and $G(f)$ the open-loop gain. We designed the unity gain frequency of $G(f)$ to not add excess fibre noise beyond $S_\nu^{SLAVE}(f)$. The yellow shaded part in Fig. 1c corresponds to $S_\nu^{NC}(f) = \left|\frac{G(f)}{1+G(f)}\right|^2 S_\nu^{fibre}(f) + \left|\frac{1}{1+G(f)}\right|^2 S_\nu^{SLAVE}(f)$, uncancelled in the synchronous comparison, determines the instability as the Dick effect limit.

**Instability of the frequency comparison**

The instability of the synchronous frequency comparison is limited by the Dick effect of non-common frequency noise $S_\nu^{NC}(f)$. The Dick effect limited stability is[19]

$$\sigma_y^2(\tau) = \frac{1}{\tau}\sum_{n=1}^{\infty}\left[\left(\frac{g_s^n}{g_0}\right)^2 + \left(\frac{g_c^n}{g_0}\right)^2\right]S_\nu^{NC}(n/T_C)/\nu_0^2,$$

where $g_0$ is the 1-cycle averaging of a sensitivity function $g(t)$, and $g_c^n$ and $g_s^n$ are the cosine and sine component of the $n$-th Fourier series expansion of $g(t)$ for a cycle time $T_C$, respectively. Purple circles in Fig. 1c shows $\sqrt{(g_c^n/g_0)^2 + (g_s^n/g_0)^2}$ for $T_C = 1.5$ s and the interrogation time $T_i = 400$ ms. Because of a sharp cut-off of $(g_c^n/g_0)^2 + (g_s^n/g_0)^2$ that decreases with $f^{-4}$ for $f > 1/T_i$, the frequency noise $S_\nu^{NC}(f)$ with $f \gg 1/T_i$ barely contributes to the Dick effect limit as shown in Fig. 1c. The Dick effect limit, therefore, may be further improved by extending $T_i$, which reduces the impact of residual fibre noise, as experimentally observed in Fig. 2e. The discrepancy between the measurements and calculations for $T_i = 400$ ms may be attributed to the frequency response of the clock excitation: although the calculation assumes a linear response to frequency, the actual Rabi excitation has a nonlinear response to frequency deviations $\delta\nu$ of a clock laser[7] unless $|\delta\nu| \ll 1/T_i$ holds, which is not the case, in particular for an extended $T_i$.

**Fibre noise cancellation with a large operational range**

The clock laser is transferred to UTokyo using a commercially available 30-km-long telecom fibre, the path of which lies along a subway train line. We found that the movement of trains shake the fibre and cause frequency deviations as large as $\Delta f = 40$ kHz with a modulation frequency of about $f_m = 200$ Hz in addition to the averaged noise and occurs every 5 minutes for a duration of 15-20 s. To cope with this frequency noise burst, we installed a VCSO (Voltage Controlled SAW Oscillator) with a frequency pulling range of 100 kHz and a digital phase and frequency detector (DPFD)[29] with a 6-bit counter in our fibre noise canceller (FNC) system, as shown in Fig. 1b. Because of the time-delay limited feedback bandwidth of $1/(4\tau_d) \approx 1.6$ kHz, the frequency deviation $\Delta f$ at $f_\text{m}$ is suppressed only down to 3 kHz which corresponds to a residual phase deviation of 30 rad. This phase excursion is temporarily stored in the 6-bit DPFD, which has a phase detection range of $64\pi$ rad, and is used to prevent cycle slips. On the other hand, in the slave side (UTokyo), in order to avoid excess fibre noise beyond the slave laser noise, we set the unity gain frequency of the loop filter to be 15 Hz as shown in Fig. 1c. Such a system allows us a robust clock link, even when the subway is in operation.

**Fibre noise suppression and extending the fibre length**

By installing the fibre noise cancellation, unsuppressed fibre delay noise[30] shows a white phase noise spectrum given by $S_\phi(f) \approx a(2\pi\tau_\text{d})^2 h_\text{L} L$ for fibre length $L$, relevant time delay $\tau_\text{d}$, and $a \sim 1$. Typical fibre phase noise is reported to be $h_\text{L} \approx 1, 4,$ and 70 (rad$^2 \cdot$ Hz)/km for Germany[24], Boulder[30], and Tokyo[6], respectively, which indicates an order of magnitude larger phase noise in Tokyo. Such an excess fibre noise makes it challenging to extend the fibre length. An installation of $(n\text{-}1)$ repeater by dividing the total fibre length by $n$ allows reducing the unsuppressed fibre noise by expanding the servo bandwidth $f_\text{BW} = n/4\tau_\text{d}$. As the fibre noise for the $L/n$ section is given by $S_\phi^n(f) \approx a\left(\frac{2\pi\tau_\text{d}}{n}\right)^2 h_\text{L} \frac{L}{n}$, fibre noise for the entire length $L$ is given by $nS_\phi^n(f) \approx a(2\pi\tau_\text{d})^2 \frac{h_\text{L}}{n^2} L$, which effectively suppresses the fibre noise by $n^2$ and improves the link stability as $\sigma_y^n(\tau) = \sigma_y(\tau)/n$.

**Tidal perturbation**

The differential gravitational attractions from extraterrestrial bodies (the moon and the sun, for example) with respect to the centre of the mass of the earth, in combination with revolution of the earth, deform the solid earth and shift of the ocean water masses, which perturb the geopotential of the earth. These phenomena are known as the solid earth tides and the ocean tides, respectively, or, collectively, as the astronomical tides. In addition, the redistributed ocean masses due to the ocean tides affect the ocean bottom as loads and induce additional deformation of the earth, which further perturb the geopotential of the earth. This phenomenon is called the ocean tide loading. Figure 4 shows tidal perturbations of the geopotential of the earth due to the astronomical tides and the ocean tidal loading, which are calculated by the software 'tide4n'[31] and 'SPOTL'[32], respectively, with minor modifications. The former uses the tidal potential developed by Tamura[33] and contains tidal constituents of periods longer than diurnal but zero-frequency, which result in DC offset in time series of tidal perturbation. The latter uses DTU10 (ref. 34) for a global ocean tide model, and NAO99Jb (ref. 35) for a regionally detailed model around Japan.

**Acknowledgements** This work is partially supported by the JSPS through its FIRST Program and from the Photon Frontier Network Program of MEXT, Japan. The authors thank K. Gibble for careful reading of the manuscript, N. Nemitz for useful comments and M. Musha for the loan of a phase locked loop circuit for the clock laser.

**Author contributions** T.T, M.T, I.U, operated clocks at UTokyo and RIKEN and analysed data. T.A, A.Y were in charge of fibre link. M.T and N.O clock lasers. Y. K. evaluated tidal perturbation. B. M. supervised spirit levelling measurements. Y. K., B. M. H. M, H.K discussed geodetic applications of a quantum altimetre. T.T, Y.K, H.K wrote the manuscript. H.K planed and supervised the experiments. All authors discussed the results and commented on the manuscript.

**Correspondence and requests for material should be addressed to H. K. (e-mail: katori@amo.t.u-tokyo.ac.jp).**

**Figures:**

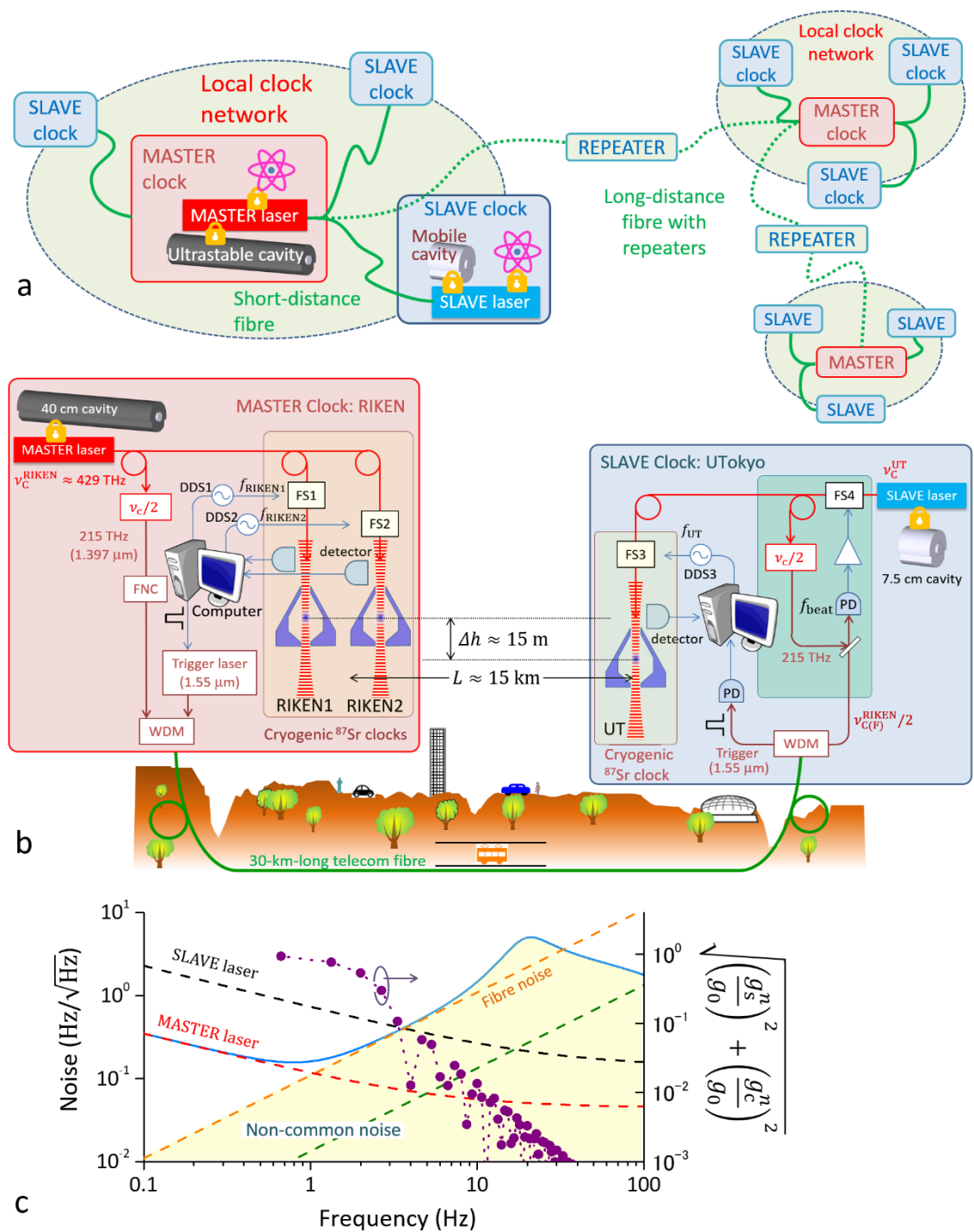

**Figure 1: Schematics of an experiment. a,** Local clock networks, each consisted of synchronously interrogated "master clock" and "slave clocks", are linked by long-distance fibres to constitute internet of clocks. **b,** Three cryogenic Sr optical lattice clocks, two (RIKEN1 and 2) at RIKEN and one (UT) at the University of Tokyo (UTokyo), act as "master" and "slave" clocks, forming a local clock network. At RIKEN, a "master laser" is stabilised to a 40-cm-long reference cavity. Its subharmonic ($\nu_\text{C}^\text{RIKEN}/2 \approx$ 215 THz) is sent to UTokyo via a 30-km-long telecom fibre equipped with a fibre noise cancellation (FNC) system to phase-lock a "slave laser" $\nu_\text{C}^\text{UT}$ pre-stabilised to a 7.5-cm-long cavity. Three clocks are steered by $f_\text{RIKEN1,2}$ and $f_\text{UT}$ generated by direct digital synthesisers (DDS1, 2, 3). The master clock sends a trigger signal to synchronously interrogate the slave clock using a 1.55-µm-laser coupled to the same fibre with wavelength-division multiplexers (WDM). PD; photodiode. **c,** Frequency noise of the master and slave lasers. The frequency noise of $\nu_\text{C}^\text{UT}$ (black dashed line) is suppressed to the blue solid line by phase-locking to $\nu_\text{C}^\text{RIKEN}/2$, which is the square root of the quadratic sum of the master laser noise $\nu_\text{C}^\text{RIKEN}$ (red dashed line) and residual fibre noise $h_\text{L}^{(\text{T})}$ (orange dashed line). Green dashed line shows the fibre noise $h_\text{L}^{(\text{G})}$ reported for the German link. Purple circles (refer to the right axis) show the sensitivity to the Dick effect, which has sharp cut off at $f_\text{c} = 1/T_\text{i}$.

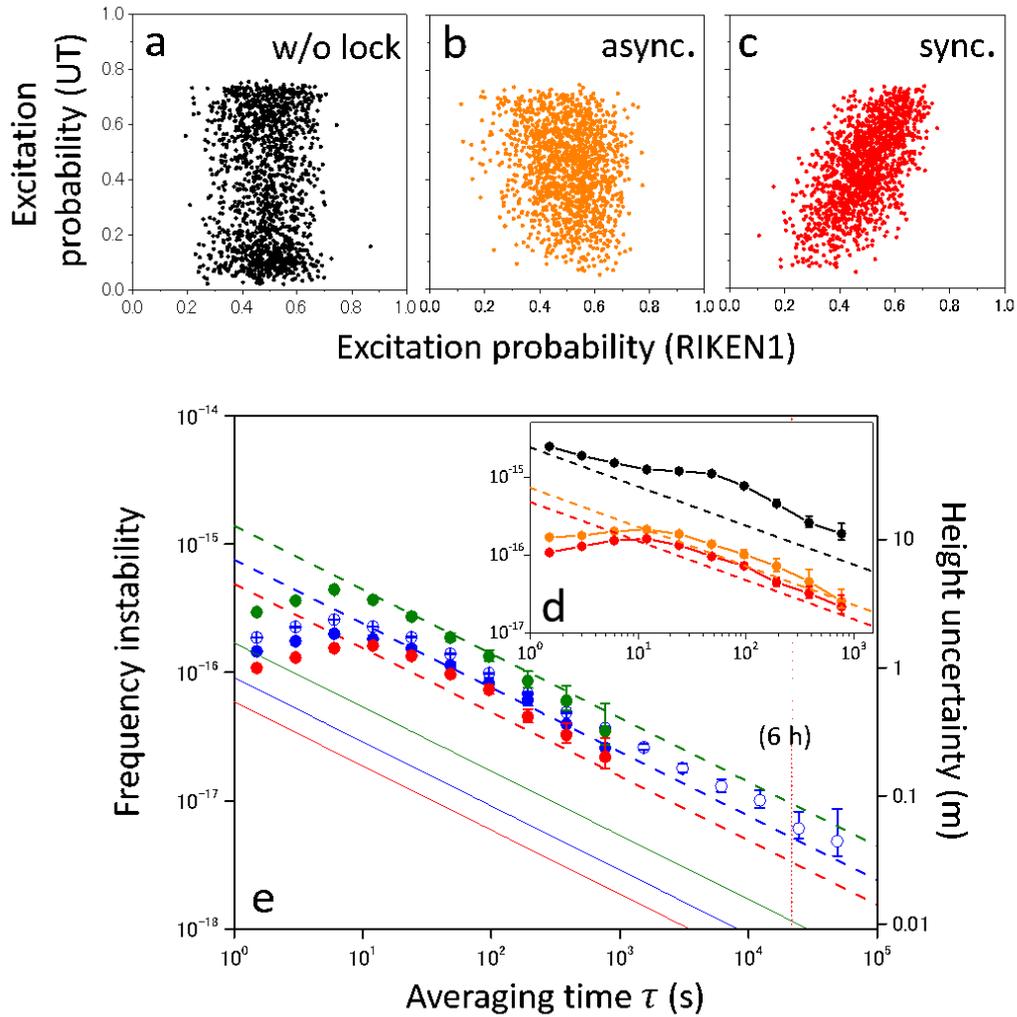

**Figure 2: Excitation probability and instability of the clock comparison. a**, Excess scatter of excitation probabilities of UT (slave clock) is reduced to **b**, by phase locking the slave laser to the master. **c**, Synchronous interrogation correlates the excitation probability. **d**, Frequency instabilities $\Delta\nu_{\text{UT}-\text{RIKEN1}}/\nu_0$ with respective colours correspond to the operation modes **a**, **b**, **c** with interrogation time $T_\text{i} = 400$ ms. **e**, Frequency instability with interrogation times $T_\text{i} = 200$ ms (green circles), 300 ms (blue circles), 400 ms (red circles), which records the lowest instability of $6 \times 10^{-16}(\tau/\text{s})^{-1/2}$. Dashed lines indicate the Dick effect limit instability for our fibre noise $h_\text{L}^{(\text{T})}$. Solid lines with corresponding colours calculated assuming the fibre noise $h_\text{L}^{(\text{G})}$ reported for the German link.

The blue empty circles show the frequency instability for the three day long measurement shown in Fig. 3b. Right axis shows corresponding height uncertainty between UTokyo and RIKEN. The error bars represent the 1σ-statistical uncertainty assuming white frequency noise.

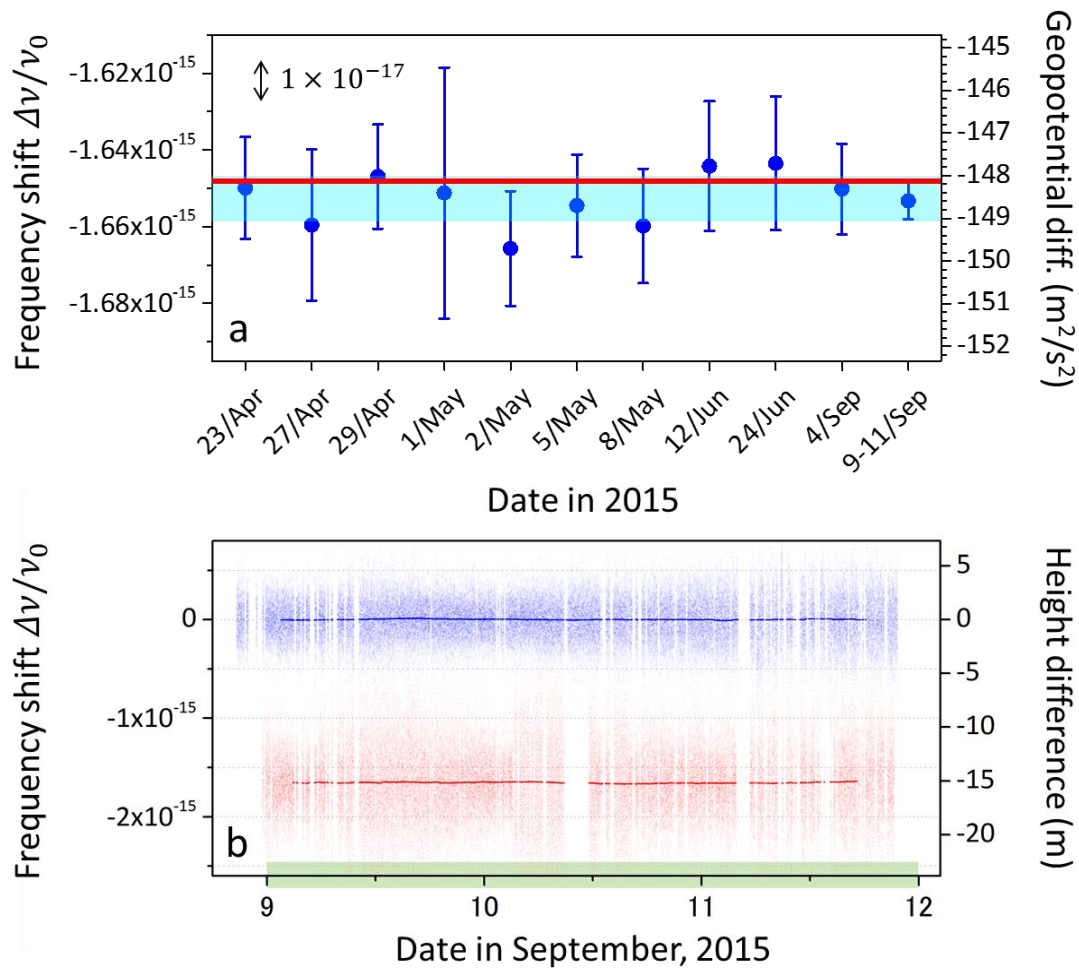

**Figure 3: Frequency difference between clocks. a,** After averaging 11 measurements (blue circles) over 6 months, we obtained the frequency difference of $\Delta\nu_{\text{UT-RIKEN1}}/\nu_0 = -1{,}652.9(5.9)\times 10^{-18}$ (blue shaded region), which is consistent with a geodetic measurement (red area). Error bars represent the 1σ-statistical uncertainty. Right axis corresponds to the geopotential difference. **b,** Frequency records of a three-day-long measurement. The "master-slave"

comparison $\Delta\nu_{UT-RIKEN1}/\nu_0$ (red dots) monitors the geopotential difference $\Delta\phi_{UT-RIKEN1}$, while the two "master" clocks at the same height (blue dots) statistically agree to $\Delta\nu_{RIKEN2-1}/\nu_0 = 1(1)\times10^{-18}$, ensuring the clocks' reproducibility. The blue and red lines show the data after 6-hour-averaging.

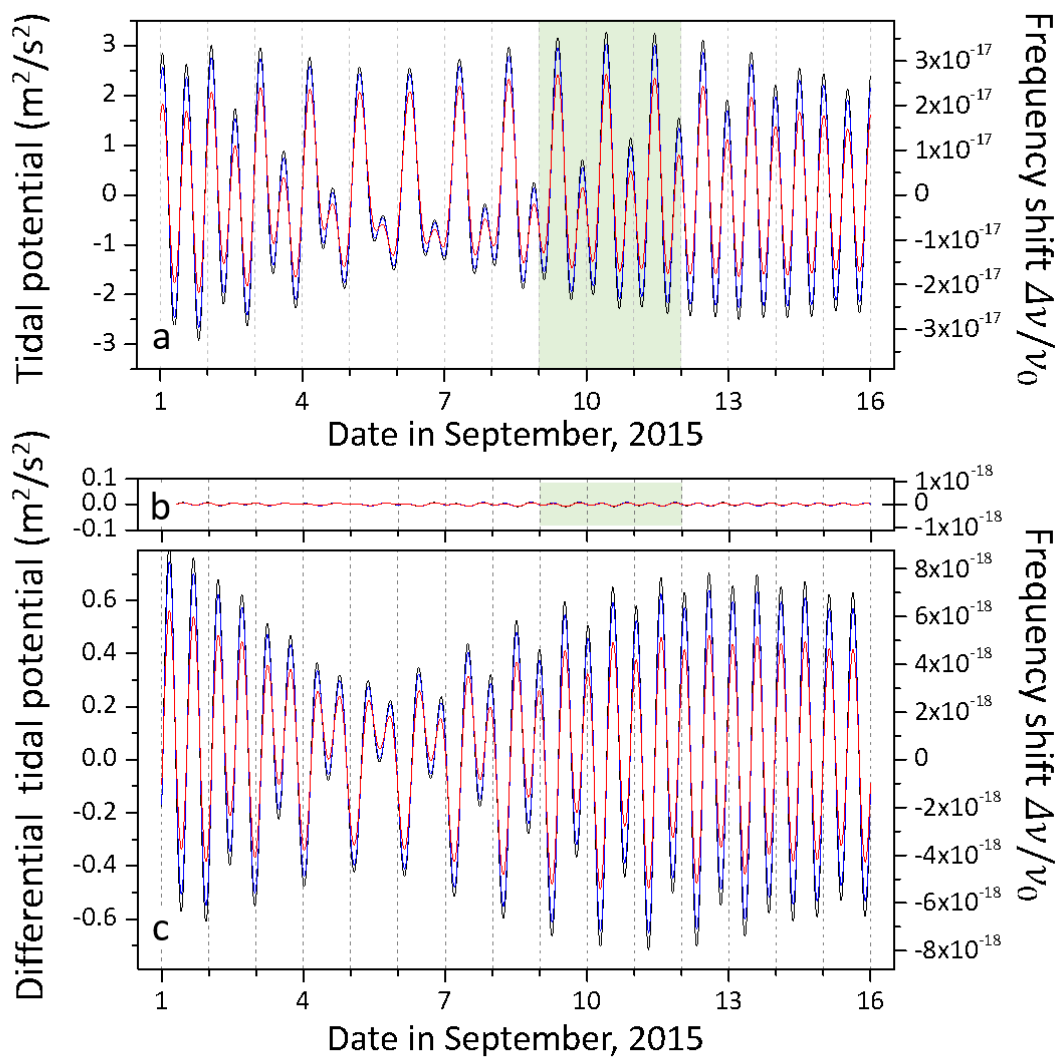

**Figure 4: Calculated tidal perturbation on geopotential for 15 days, from September 1, 2015. a,** Astronomical tides (body tides) and ocean tidal loading are simulated and combined to form the tidal perturbation at RIKEN. To compare to the frequency differences of the separated clocks, the perturbation

differences are calculated, **b,** between UTokyo and RIKEN, and, **c,** between Akune and RIKEN of about 1,000 km apart. The perturbation is calculated every 7.5 minutes (black curve), and running averages are computed with a 6-hour (red curve) and a 3-hour window (blue curve). The green shaded region corresponds to the measurements in Fig. 3b.

**Table 1: Uncertainty budgets for clocks.** The systematic corrections for the UT clock ($\nu_{\mathrm{UT}}$) and for the fractional frequency difference $\Delta\nu_{\mathrm{UT-RIKEN1}} = \nu_{\mathrm{UT}} - \nu_{\mathrm{RIKEN1}}$ are shown. Representative values for the three-day-long campaign (see Fig. 3b) are given. *The cryogenic chamber for UT (RIKEN1) operates at a temperature of 105 K (95 K). †See Methods.

| Contributor | $^{87}$Sr (UT) | | $^{87}$Sr (UT)-$^{87}$Sr (RIKEN1) | |
|---|---|---|---|---|
| | Correction ($10^{-18}$) | Uncertainty ($10^{-18}$) | Correction ($10^{-18}$) | Uncertainty ($10^{-18}$) |
| Quadratic Zeeman shift | 109.0 | 0.9 | -8.2 | 0.5 |
| Blackbody radiation shift* | 79.1 | 0.9 | 24.9 | 1.2 |
| Lattice light shift† | 3.5 | 5.0 | 0.0 | 4.4 |
| Clock light shift | 0.047 | 0.023 | 0.0 | 0.014 |
| First-order Doppler shift | 0.0 | 0.5 | 0.0 | 0.7 |
| AOM chirp & switching | 0.0 | 0.2 | 0.0 | 0.3 |
| Servo error† | 1.3 | 3.9 | 0.6 | 0.5 |
| Density shift† | 1.1 | 5.2 | 0.7 | 3.3 |
| Systematic total | 194.0 | 8.3 | 18.0 | 5.7 |